\documentclass[reprint, 
superscriptaddress,
 amsmath,amssymb,
 aps, physrev,
 pra,
showkeys
]{revtex4-2}

\usepackage{graphicx}
\usepackage{dcolumn}
\usepackage{bm}
\usepackage{mathrsfs}
\usepackage{hyperref}

\usepackage{tikz}
\usepackage{lipsum}

\begin{document}


\title{\textbf{Quantum correlations in entangled two-particle non-unitary quantum walks} 
}%

\author{Gene M. M. Itable}
\email{gmmitable@gmail.com}
\affiliation{
Department of Science and Technology - Advanced Science and Technology Institute, Quezon City, Philippines
}

\author{Francis N. C. Paraan}
\email{fparaan@nip.upd.edu.ph}
\affiliation{
National Institute of Physics, University of the Philippines Diliman, Quezon City, Philippines}

\date{\today}

\begin{abstract}
    We study the evolution of quantum correlations in two-particle discrete-time non-unitary quantum walks on a line with gain and loss. The two particles are initially prepared in a maximally entangled state and evolve independently. Using numerically exact calculations of position probability densities, average interparticle distance, entanglement entropy, and concurrence, we examine how dissipation and particle-exchange symmetry of the state (symmetric or antisymmetric) influence the resulting correlations. Notably, we find that the entanglement entropy of the antisymmetric state decays slower than that of the symmetric state at the exceptional point of the model. We also show that, in the antisymmetric state, the concurrence measured on the one-particle reduced density operator is unaffected by dissipation. Furthermore, we discuss how these correlations indicate quantum-to-classical transitions through the appearance of Gaussian position probability densities, sub-linear growth of average interparticle distance after a few steps, and decay of the entanglement entropy. On the other hand, the robustness of the qubit entanglement entropy can differentiate the model's topological phases better than in the one-particle quantum walk scenario.
\end{abstract}

\keywords{Quantum walks, non-unitary dynamics, dissipation}
\maketitle

\section{Introduction} 

The effects of uncontrolled interactions between a quantum system and the environment pose challenges in realizing practical processing of quantum information for computing, sensing, and communication applications \cite{aolita2015open,suter2016colloquium,harrington2022engineered}. These uncontrolled interactions may manifest in different mechanisms, such as dissipation of the energy of a quantum system which then induces decoherence. There are various ways to incorporate dissipation in models of open quantum systems, one of which is with effective non-Hermitian Hamiltonians, or equivalently, with non-unitary dynamics. Hence, there has been considerable interest to characterize quantum information in non-Hermitian models and to develop processes with non-Hermitian models \cite{croke2015pt,mochizuki2016explicit,sergi2016quantum,izaac2017quantum,ju2019non,fring2019eternal,gopalakrishnan2021entanglement, itable2023entanglement,badhani2024non,zou2022bell}.

Quantum correlations, such as entanglement, are valuable resources in quantum information processing. Their preservation against dissipation or their utilization despite dissipation is important \cite{suter2016colloquium,harrington2022engineered,aolita2015open}. 
As a resource for some quantum algorithms, a highly entangled state encodes the initial quantum information, with which its entanglement decreases as quantum information is processed \cite{chitambar2019quantum}. While dissipation may be detrimental to entanglement \cite{aolita2015open,yu2009sudden}, some non-Hermitian models with combined parity and time-reversal symmetries (or PT-symmetry) have shown non-vanishing entanglement in some regions of its parameter space \cite{chakraborty2019delayed,fring2019eternal,gopalakrishnan2021entanglement}. 

A popular framework to model quantum information processes is with quantum walks; including those that incorporate dissipation \cite{attal2012open,attal2012open_graphs,sinayskiy2012efficiency,mochizuki2016explicit,itable2023entanglement,badhani2024non}. Generally, a quantum walk models the spreading of a quantum state on graphs \cite{venegas2012quantum,wang2013physical,kadian2021quantum}. 
Dissipation in quantum walks has been examined from various perspectives in the literature, often to describe quantum-to-classical transitions. An explicit approach is to model the environment as a bath coupled to the quantum walk. In a general bosonic bath with the quantum walk dynamics encoded as a quantum channel, the quantum walk reduces to a classical random walk in the long-time limit \cite{attal2012open,attal2012open_graphs,sinayskiy2019open}. In a thermal phonon bath in which the master equation is derived for the quantum walk and a dissipative parameter is defined, the quantum walk reduces to a classical random walk as the dissipation parameter is increased \cite{nizama2012non,nizama2014quantum}. This was observed through various quantities that characterize quantum correlations, i.e. position probability distribution, entanglement entropy, concurrence, negativity, and discord.

Another approach is through non-Hermitian quantum walks. Here, the environment-induced energy gain and loss on the quantum walk are incorporated effectively in the Hamiltonian or in the non-unitary evolution operator. Non-Hermitian quantum walks on the line have been explored on a model with a decay-inducing sublattice \cite{rudner2009topological} and on split-step models with alternating gain and loss \cite{regensburger2012parity,mochizuki2016explicit, mittal2021persistence}. These models elucidate the nature of topological phase transitions in non-Hermitian quantum walks. Experiments in photonic setups have supported the presence of topological phases through observations of topological edge states \cite{xiao2017observation,wang2020observation} and non-Hermitian bulk-boundary correspondence \cite{xiao2020non}, and have shown how the spectral topology affects the dynamics \cite{xue2024self}. Robust entanglement can be an indicator of the topological phases in non-Hermitian quantum walks \cite{itable2023entanglement}, in a similar way as in Hermitian quantum walks \cite{kitagawa2010exploring,wang2020robustness}. When dissipation is strong enough, the spatial distribution becomes similar to a classical random walk \cite{mochizuki2016explicit} and the entanglement decays \cite{itable2023entanglement}. As an algorithmic framework, PT-symmetric quantum walks have been applied in centrality testing which showed an advantage over classical algorithms in some cases \cite{izaac2017quantum}.

One-particle quantum walks may offer an advantage on some tasks over their classical random walk counterparts and for a thorough review see Ref. \cite{venegas2012quantum, wang2013physical, kadian2021quantum}. However, since one-particle quantum walks can be simulated with classical resources \cite{knight2003optical,jeong2004simulation}, multi-particle quantum walks can offer quantum resources that are truly unavailable classically, e.g. entanglement between the particles. In random walks, non-interacting particles will not form correlations with one another and the joint position probability density is simply the product of one-particle position probability distribution. On the other hand, quantum walks even without particle interactions can generate spatial correlations that depend on the symmetry of the initially entangled state under particle exchange \cite{omar2006quantum, vstefavnak2011directional, rigovacca2016two, muhammad2024spatial}.

For example, in two-particle unitary quantum walks that is initially symmetric (antisymmetric) under particle exchange, the average distance can be smaller (larger) than if they are initially in a product state \cite{omar2006quantum}. Furthermore, the initially two-particle symmetric state tends to be at the same side of the line which can be increased by introducing an interaction that breaks the translation symmetry \cite{vstefavnak2011directional}. 
These reflect the fact that symmetric two-particle states may represent bosons that tend to bunch together. On the other hand, antisymmetric states may represent fermions that tend to keep their distance from one another (anti-bunch). Initial entanglement between the two particles can also be preserved if the initial and final states have opposite exchange symmetries after post-selected measurements \cite{muhammad2024spatial}.  

Decoherence on two-particle quantum walks has been explored through percolated lattices with randomly generated missing links on the line \cite{rigovacca2016two}. The work shows how the spatial correlations depend on the particle-exchange symmetry of the states even at low levels of percolation. However, decoherence due to dissipation in two-particle quantum walks remains unexplored. In this paper, we study the case of two particles in a non-unitary split-step quantum walk with alternating gain and loss. We examine the spatial correlations, the entanglement entropy, and the concurrences between all unique biparitions to understand the role of the particle-exchange symmetry in the initial state and the effects of dissipation in these non-unitary quantum walks.

This paper is organized as follows. In Section \ref{sec:sec1} we provide a background on a model of one-particle non-unitary quantum walks with balanced gain and loss. We then describe the non-interacting two-particle non-unitary quantum walk version. In Section \ref{sec:sec3}, we discuss how the initial symmetry under the exchange of the two particles leads to distinct spatial correlations. In Section \ref{sec:sec4}, we discuss the entanglement entropy dynamics and the presence of genuine multipartite entanglement through the concurrence dynamics. We summarize our findings in Section \ref{sec:sec5} and provide our recommendations for future work.

\section{Non-unitary quantum walks} \label{sec:sec1}

\subsection{Translation-invariant one-particle quantum walks}

The Hilbert space \( \mathcal{H} \) for a one-particle quantum walk on the line is \( \mathcal{H}=\mathcal{H}_\text{p} \otimes \mathcal{H}_\text{q} \) where \( \mathcal{H}_p = \text{Span}(\{ |n\rangle \; | \; n \in \mathbb{Z}  \}) \) represents the localized position states along the lattice and \( \mathcal{H}_q = \text{Span}(\{ |0\rangle, |1\rangle \}) \) represents the internal two-level or qubit states of the particle/walker. The general state in this Hilbert space \(H \) can be written as \( | \psi(t) \rangle = \sum_{n,q} \psi_{n,q}(t) |n\rangle \otimes |q\rangle \) such that \(n \in \mathbb{Z} \) and \(q=0,1\). For brevity, we write composite states \(|n\rangle \otimes |q\rangle \) as \( |n,q \rangle\). 

A general quantum walk model on a one-dimensional lattice consists of a shift operator \(S \) conditioned on the qubit state of the particle,
\begin{equation}
    \mathcal{S} = \sum_n \big( |n-1\rangle \langle n| \otimes |0\rangle \langle 0| +  |n+1 \rangle \langle n| \otimes |1\rangle \langle 1| \big),
\end{equation}
and an operator \(\mathcal{C} = \sum_n |n\rangle\langle n| \otimes C(\theta_n) \) that transforms the qubit state at site \(n \) through the local operator \(C(\theta_n) \) with parameter \(\theta_n \). In most works, the operator \(C \) is called the coin operator with an action that is analogous to the flipping of a coin in a classical one-dimensional random walk which decides the direction of the next step. The evolution of the particle per time step is governed by \(\mathcal{U} = \mathcal{S}\mathcal{C} \). Hence, the state of the particle after \(t \) steps is given by \(|\psi(t)\rangle = \mathcal{U}^t |\psi(0)\rangle \) for some initial state \(|\psi(0)\rangle \).

A translation-invariant model may be defined with site-independent operator \(\mathcal{C} = \mathbb{I}_\text{p} \otimes C(\theta) \) which leads to the evolution governed by \(\mathcal{U}=\mathcal{S}(\mathbb{I}_\text{p} \otimes C(\theta)) \). In practice, simulating the quantum walk with \(T \) steps allows us to truncate the position Hilbert space \(\mathcal{H}_\text{p} \) to \(\mathcal{H}^T_\text{p} = \text{Span}(|-T\rangle,...,|0\rangle,...,|T\rangle) \), an \(N\)-dimensional Hilbert space with \(N=2T+1 \). Due to translational invariance, it helps to use a Fourier basis set \(\{ |k\rangle \} \) for \(\mathcal{H}^T_\text{p} \) using discrete Fourier transform. The basis states are labeled by wavevectors \(k \) in first Brillouin zone, \(k \in \text{BZ} = \{-\pi+\kappa \Delta k \; | \; \kappa = 0,...,N-1 \} \) and \(\Delta k = 2\pi/N \). These sets of basis are related to another by \(|k\rangle =\sum_n e^{\mathrm{i} k n} |n\rangle / \sqrt{N} \) and \(|n\rangle = \sum_k e^{-\mathrm{i} k n} |k\rangle / \sqrt{N} \). The general state of the particle can be written as
\begin{equation}
    |\psi(t) \rangle =   \sum_k |k,\tilde{\psi}_k(t) \rangle
\end{equation}
where \( |\tilde{\psi}_k(t)\rangle \) is the particle's qubit state per mode \(k \). In this basis, \(\mathcal{U} \) becomes
\begin{equation}
    \mathcal{U} = \sum_k |k\rangle \langle k| \otimes U_k, \quad U_k=S_k C(\theta),
\end{equation}
where the shift operator becomes \(S_k = e^{\mathrm{i} k\sigma_3} \). The evolution of the system is simply obtained from the modes with their evolutions given by
\begin{equation}
     |\tilde{\psi}_k(t)\rangle = (U_k)^t  |\tilde{\psi}_k(0)\rangle,
\end{equation}
where \(|\tilde{\psi}_k(0)\rangle \) are their initial states. This formalism simplifies the simulation of the system and the calculation of some relevant properties specially those that would correspond to measuring the state of the qubit. For example, the reduced density operator for the qubit is simply
\begin{equation}
    \rho_\text{q}(t) = \text{Tr}_\text{p} (|\psi(t)\rangle \langle \psi(t)| ) =  \sum_k |\tilde{\psi}_k(t)\rangle \langle \tilde{\psi}_k(t)|.
\end{equation}
For a particle initialized at the origin and in qubit state \(|q\rangle \), i.e. \( |\psi(0) \rangle = |0,q\rangle \), all the modes are initially set in \(|\tilde{\psi}(k,0)\rangle=|q\rangle \). When considering the infinite-time limit of the quantum walk in which the truncated line becomes infinite we need to only convert the summation into integration with \(N^{-1/2} \sum_k \rightarrow (2\pi)^{-1/2} \int \mathrm{d}k \) since \(\Delta k = 2\pi/N \).

\subsection{One-particle non-unitary quantum walks}

Non-Hermitian Hamiltonians with combined parity and time-reversal symmetries or PT-symmetry have gained significant attention since despite non-Hermiticity, they may possess real eigenvalues that would correspond to real energies \cite{ashida2020non}. A Hamiltonian \( H \) has PT-symmetry when it is unchanged under the combined action of the reflection about the origin, \( \mathcal{P}: (\vec{x} \rightarrow -\vec{x}) \), \( \vec{x}\) here being the coordinates, and the reversal of time \(t\), \( \mathcal{T}: (t \rightarrow -t) \), i.e. \([H,\mathcal{P}\mathcal{T}] =0\). When the eigenvectors of the PT-symmetric non-Hamiltonian are also PT-symmetric, then the eigenvalues are real. Otherwise, the eigenvalues are complex. Another unique property of non-Hermitian Hamiltonians is the existence of exceptional points in which eigenvalues and their respective eigenvectors coalesce. 

We consider a model of non-unitary quantum walks \cite{mochizuki2016explicit} that was inspired by experiments in photonic lattices with PT-symmetry \cite{regensburger2012parity}. In the Fourier basis, the evolution is given by
\begin{equation}
    U_k = S_k G(\gamma) \Phi(\phi) C(\theta_2) S_k G^{-1}(\gamma) \Phi(\phi) C(\theta_1), \label{eq:gainloss_model}
\end{equation}
the \(C(\theta_i) = e^{\mathrm{i} \theta_i \sigma_1} \) is the coin operator, 
\( G(\gamma) = e^{\gamma \sigma_3} \) is the gain-loss operator, and \(\Phi = e^{\mathrm{i} \phi \sigma_3} \) is the phase operator. \(\{  \mathbb{I}_\text{q}, \sigma_1,\sigma_2,\sigma_3\} \) are the Pauli operators . The model parameters \(\theta_i,\phi \) and \(\gamma \) are real. Hence, the quantum walk is unitary when the gain-loss parameter is \(\gamma=0\) or \(e^\gamma = 1\), and non-unitary otherwise. This type of quantum walk is sometimes called a split-step quantum walk in which a single time step is split into two consecutive shift operations. Since the particle will only have nonzero amplitudes at even sites, we can trunctate the position Hilbert space to contain only the even sites. The appropriate Brillouin zone for this case is \([-\pi/2,\pi/2) \) due to the two units of lattice spacing. 

The model here has been studied thoroughly in Ref. \cite{mochizuki2016explicit}. The quasi-energies of the effective Hamiltonian \( H_k \) from \( U_k = e^{-\mathrm{i} H_k} \) is given by \( \pm \epsilon_k = \arccos\big(\cos \theta_1 \cos \theta_2 \cos \big(2(k+\phi)\big)  - \sin \theta_1 \sin \theta_2 \cosh (2\gamma) \big) \). Exceptional points occur when the quasi-energy band gap closes at certain $k$. The unbroken and broken PT-symmetry phases can be distinguished based on their energy spectra. The former has a gapped and entirely real energy spectrum, while the latter has imaginary effective energies for some modes. These two phases are separated by exceptional points, where the energy bands remain real but develop non-analytic cusps which marks the PT-symmetry breaking phase transition. In photonic experimental realizations of this and similar quantum walks, beam displacers typically implement the shift operators, partially polarizing beam splitters implement the gain-loss operators, and wave plates implement phase shifts and general coin rotations \cite{regensburger2012parity,xiao2017observation,wang2020robustness,mochizuki2020bulk}.

The non-unitarity of the model implies an evolution of the state that does not preserve the total probability. This makes the standard density operator \(\rho(t) = |\psi(t)\rangle \langle \psi(t)| \) non-Hermitian. In this study, we adopt the framework of normalizing at each time step to obtain a Hermitian density operator,
\begin{equation}
     \varrho(t) \equiv \frac{\rho(t)}{\text{Tr}\big(\rho(t) \big)} = \frac{|\psi(t)\rangle \langle \psi(t)| }{\text{Tr} \big(|\psi(t)\rangle \langle \psi(t)| \big)},
\end{equation}
which is the conventional framework when non-Hermitian systems are considered as effective models of dissipative dynamics with no abrupt transitions or quantum jumps \cite{herviou2019entanglement,lee2014heralded}. An instantaneously normalized density operator \( \varrho(t) \) guarantees that the reduced density operator spectrum, the entanglement spectrum, corresponds to a probability distribution as has been shown in the study of entanglement in fermionic chains \cite{herviou2019entanglement}, magnetic models \cite{lee2014heralded}, and in other non-Hermitian qudit models \cite{sergi2016quantum, wen2021stable}. Its use can be further justified since the raw intensity \( \langle {\psi(t)} | {\psi(t)} \rangle\) matches with the observed amplified photon counts in the broken PT-symmetry phase of the corresponding experimental system \cite{regensburger2012parity, xiao2017observation, xiao2020non}. 

We must note that there are other approach that constructs density operators given a non-unitary evolution, such as a metric formulation of inner products that satisfies a number of no-go theorems for quantum information \citep{ju2019non}. Badhani et al. characterize how distinguishability between distinct initial states is lost due to the non-unitarity of the model (Eq. (\ref{eq:gainloss_model})) through an information back-flow measure under an instantaneously normalized density operator approach and a metric formalism \cite{badhani2024non}. They showed that the two methods produce the same information back-flow measure in the unbroken PT-symmetry phase, but then the measure has numerical deviation, although with similar qualitative features, near the exceptional point and in the broken PT-symmetry phase. 

\subsection{Two-particle non-unitary quantum walks}

 The full Hilbert space for the two-particle quantum walk on the line is \(\mathcal{H}=\mathcal{H}_{\text{S}_1} \otimes \mathcal{H}_{\text{S}_2} \) such that \(\mathcal{H}_{\text{S}_i} = \mathcal{H}_{\text{p}_i} \otimes \mathcal{H}_{\text{q}_i}\). We denote the two-particle position subspace as \(\mathcal{H}_\text{P} = \mathcal{H}_{\text{p}_1} \otimes \mathcal{H}_{\text{p}_2} \), while the internal two-qubit subspace as \(\mathcal{H}_\text{Q} = \mathcal{H}_{\text{q}_1} \otimes \mathcal{H}_{\text{q}_2}  \). Hence, we have two sets of Fourier basis for the two particles \(\{ |k_1\rangle \} \) and \(\{ |k_2\rangle \} \). The general state of the system can be written as
\begin{equation}
    |\psi(t)\rangle = \sum_{k_1,k_2}  |k_1,\tilde{\psi}_{k_1}(t)\rangle \otimes |k_2,\tilde{\psi}_{k_2}(t)\rangle.
\end{equation}
We consider two non-interacting indistinguishable particles for this quantum walk, the evolution operator is simply \(\mathcal{U} \otimes \mathcal{U} \). This translates to separate evolutions per particle and per mode, i.e.
\begin{equation}
    | \tilde{\psi}_{k_i}(t) \rangle = (U_{k_i})^t  | \tilde{\psi}_{k_i}(0) \rangle, \quad i=1,2.
\end{equation}

We consider the two maximally entangled states,
\begin{equation}
    | \psi_\pm \rangle = \frac{1}{\sqrt{2}} \big( |0,0\rangle \otimes |0,1\rangle \pm |0,1\rangle \otimes |0,0\rangle \big),\label{eq:symmetric_antisymmetric_states}
\end{equation}
as the initial states of the quantum walk. With respect to the exchange symmetry, \(|\psi_+ \rangle \) is symmetric, while \(|\psi_-\rangle \) is antisymmetric. As the two particles are indistinguishable, the symmetric state may represent a quantum walk with two bosons, while the antisymmetric state may represent a quantum walk with two fermions.  

The general state of two indistinguishable particles is \(|\psi\rangle = \sum_{m,n} c_{m,n} |m\rangle \otimes |n\rangle \) where \( \{ |m\rangle \} \) spans the one-particle Hilbert space. Let \(\Pi \) be the particle exchange operator, i.e. \(\Pi( |m\rangle \otimes |n\rangle) = |n\rangle \otimes |m\rangle \). Consider a linear operator \( \mathcal{U}_{1,2} \) on the two-particle Hilbert space. For \(\mathcal{U}_{1,2}\) to commute with \(\Pi\),  \(\mathcal{U}_{1,2}\) must satisfy
\begin{equation}
    \mathcal{U}_{1,2} (|n\rangle \otimes |m\rangle) = \Pi \; \mathcal{U}_{1,2} (|m\rangle \otimes |n\rangle ).
\end{equation}
Clearly, identical local linear operations on the two particles , i.e. \(\mathcal{U}_{1,2} = \mathcal{U} \otimes \mathcal{U}\), will satisfy the above condition. Since the evolution operator \( \mathcal{U}\otimes\mathcal{U} \) commutes with the exchange operator \( \Pi \), then the the symmetry or antisymmetry of the initial state is preserved throughout the quantum walk. 

In the Fourier basis, these states \(|\psi_\pm\rangle\) can be written as
\begin{equation}
    |\psi_{\pm} \rangle = \frac{1}{\sqrt{2}}\sum_{k_1,k_2} \big( |k_1,0\rangle \otimes |k_2,1\rangle \pm |k_1,1\rangle \otimes |k_2,0\rangle \big). 
\end{equation}
Let us define \( | \tilde{\psi}^{q_1}_{k_i}(t) \rangle = (U_{k_i})^t |q_i \rangle \) such that \( q_i = 0,1 \) representing the initial qubit state, and \(\rho^{q_1,q_2}_{k_i,k_i'}(t) = |\tilde{\psi}^{q_1}_{k_i}(t)\rangle \langle \tilde{\psi}^{q_2}_{k_i'}(t)| \). The time evolution of the density operator can be written as
\begin{widetext}
    \begin{equation}
        \rho^\pm(t) = \frac{1}{2} \sum_{q,q'} (\pm1)^{q\oplus q'} \bigg( \sum_{k_1,k_1'} |k_1\rangle \langle k_1'| \otimes \rho_{k_1,k_1'}^{q,q'}(t) \bigg) \otimes \bigg( \sum_{k_2,k_2'} |k_2\rangle \langle k_2'| \otimes \rho_{k_2,k_2'}^{\bar{q},\bar{q}'}(t) \bigg), \label{eq:explicit_density_operator}
    \end{equation}
where \(q\oplus q' \) is the addition modulo 2 and that \(\bar{q} = q \oplus 1\). For example, the reduced density operator on the two-particle position subspace is simply
    \begin{equation}
        \rho^\pm_\text{P}(t) = \frac{1}{2} \sum_{q,q'} (\pm1)^{q\oplus q'} \bigg( \sum_{k_1,k_1'} |k_1\rangle \langle k_1'| \; \text{Tr}\big( \rho_{k_1,k_1'}^{q,q'} (t) \big) \bigg) \otimes \bigg( \sum_{k_2,k_2'} |k_2\rangle \langle k_2'| \;\text{Tr}\big(\rho_{k_2,k_2'}^{\bar{q},\bar{q}'}(t) \big) \bigg). \label{eq:reduced_density_operator_2position}
    \end{equation}
\end{widetext}
The normalized density operator follows as \(\varrho^\pm(t) = \rho^\pm(t) / \text{Tr} \big( \rho^\pm(t) \big) \). If the initial state \(|\psi(0)\rangle = |0,0\rangle \otimes |0,1\rangle \), we can immediately see that the resulting density operator is simply \(\rho \sim  \sum |k_1\rangle \langle k_1'| \otimes \rho^{0,0}_{k_1,k_1'}(t) \otimes  |k_2\rangle \langle k_2' | \otimes \rho^{1,1}_{k_2,k_2'}(t) \). As expected, the two particles can be separately described by their own density operators. Hence, initializing the particles in an entangled state will generate correlations between them through the quantum walk without explicit interparticle interaction. 

\section{Spatial correlations} \label{sec:sec3}

\begin{figure*}
    \centering
    \includegraphics[width=\linewidth]{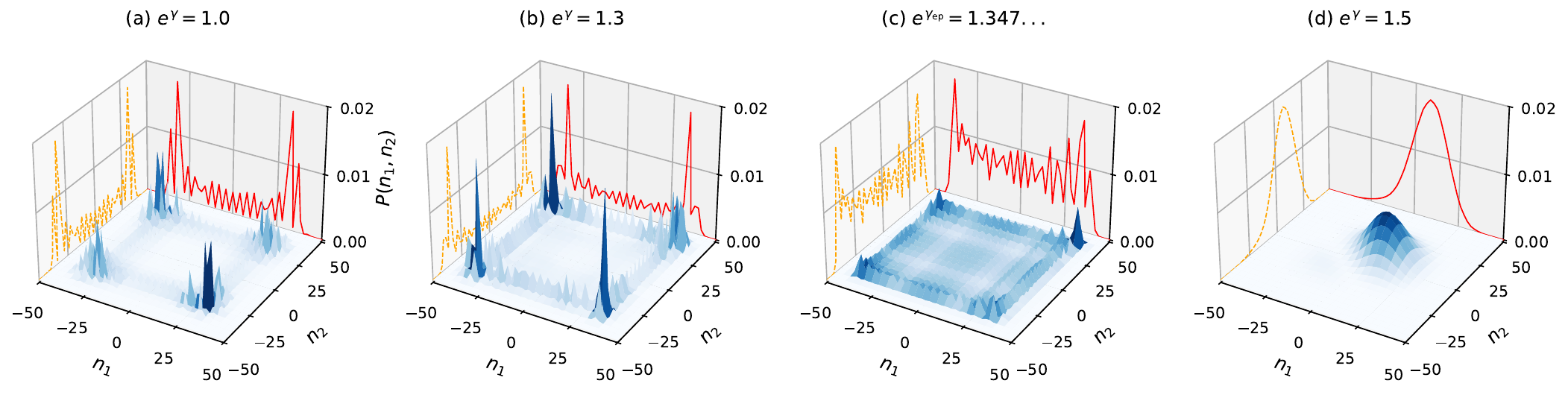}\\
    \includegraphics[width=\linewidth]{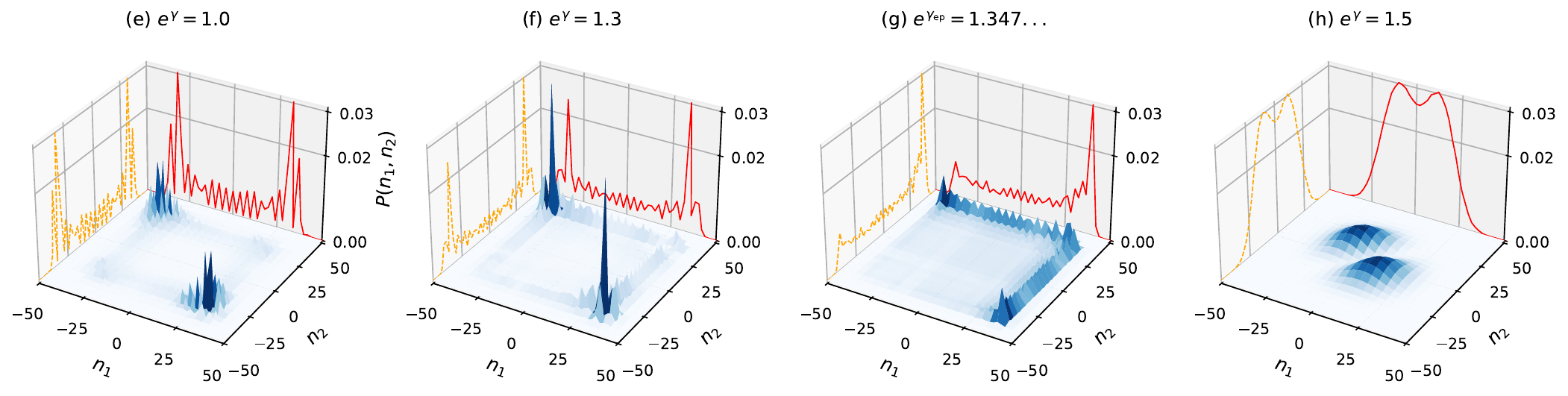}\\
    \caption{Joint position probability densities \(P(n_1,n_2) \) at time step \(t=25 \) for increasing gain-loss parameter \( \gamma \) with the other model parameters set at \(\theta_1=\pi/4, \theta_2 =-\pi/7\) and \(\phi=0\). The top row (a)-(d) corresponds to the non-unitary quantum walks initially in the symmetric state \(|\psi_+\rangle \), while the bottom row (e)-(h) corresponds to the antisymmetric state \(|\psi_-\rangle \). For \( e^\gamma <  e^{\gamma_\text{ep}} \) the quantum walk has unbroken PT-symmetry, large \(P(n_1=\pm 2t,n_2=n_1) \) indicates bunching of the two particles for the symmetric state \(|\psi_+ \rangle \), and large \(P(n_1=\pm 2t,n_2=-n_1) \) indicates anti-bunching for the antisymmetric state \( |\psi_-\rangle \). At the exceptional point \(e^{\gamma_\text{ep}}\), the particles can be measured at any pair of sites with almost similar probability for \( |\psi_+\rangle\), while anti-bunching still manifests in the case of \(|\psi_-\rangle \) with one particle definitely at \(n_i=\pm 2t\) and the other at \(n_j \neq n_i \). Lastly, for \( e^\gamma >  e^{\gamma_\text{ep}} \) the quantum walk has broken PT-symmetry and \(P(n_1,n_2) \) can be approximated by Gaussian distribution which implies transition to a classical random walk. The peak of the \(P(n_1,n_2) \) still indicates bunching or anti-bunching of the two particles. The one-particle position probability densities are shown in the vertical planes; \(P(n_1)\) in solid red line and \(P(n_2)\) in dashed orange lines, they are not drawn to scale.}
    \label{fig1}
\end{figure*}

The evolution operator \(\mathcal{U}=U\otimes U \) we consider is separable with respect to the Hilbert space of each of the particle. The evolution will not produce correlations between the particles if the initial state is separable \( |\psi(0)\rangle = |\psi_1\rangle |\psi_2\rangle \). Initializing the particles in a symmetric and antisymmetric entangled states, in Eq. (\ref{eq:symmetric_antisymmetric_states}), lead to distinct spatial correlations in unitary quantum walks \cite{omar2006quantum,vstefavnak2011directional}. In this non-unitary quantum walk model, it is interesting to see the effects of the gain-loss operation. Spatial properties can be obtained from the reduced density operator on the two-particle position subspace, \( \rho^\pm_\text{P}(t) = \text{Tr}_\text{Q}\big(\rho^\pm(t)\big) \).

\subsection{Joint position probability densities}

We consider the joint position probability density,
\begin{equation}
    P(n_1,n_2,t) = \langle n_1,n_2| \varrho_{\text{P}}(t) |n_1,n_2\rangle,
\end{equation}
such that \(\rho_\text{P}(t)\) is given in Eq. (\ref{eq:reduced_density_operator_2position}), to examine the spatial correlations at different gain-loss parameter \(\gamma \) for the symmetric and antisymmetric initial states, \( |\psi_\pm \rangle \). Fig. \ref{fig1} shows the joint position probability density \( P(n_1,n_2) \) for \(t = 25 \) steps with model parameters set at \( \theta_1 = \pi/4,\theta_2 =-\pi/7 \), and \(\phi=0 \). Below the exceptional point \( e^\gamma < e^{\gamma_\text{ep}} \) in which \( e^\gamma = 1.0 \) represents the unitary limit (Fig. \ref{fig1} (a) and (e)) and \( e^\gamma = 1.3 \) represents a non-unitary quantum walk in the unbroken PT-symmetry phase (Fig. \ref{fig1} (b) and (f)) , we observe the same features consistent to unitary two-particle quantum walks \cite{omar2006quantum}. The joint position probability density \(P(n_1,n_2)\) shows non-zero probability of bunching, i.e. the two particles being measured at \(n_1=n_2=\pm 2t\), for the symmetric state \( |\psi_+ \rangle \). On the other hand, there is almost zero probability of bunching for the antisymmetric state \( |\psi_-\rangle \), and hence, there is significant probability of anti-bunching, the two particles being measured at \(n_1=\pm 2t,n_1 \neq n_2 \). The result agrees to the bunching (anti-bunching) behavior expected from bosons (fermions) which has to be described by a symmetric (antisymmetric) state.

When the gain-loss parameter is large \(e^\gamma > e^{\gamma_\text{ep}} \) that the evolution breaks the PT-symmetry (in Fig. \ref{fig1} (d) and (e)), the joint position probability density \(P(n_1,n_2)\) can be approximated by Gaussian distribution. This indicate that the non-unitary quantum walk can also be described by a classical random walk \cite{nizama2012non}. However, we see in the antisymmetric initial state \( |\psi_-\rangle \) a strong spatial anti-correlation. This is also observed in a unitary two-particle quantum walk with non-interacting particles with the two-qubit initially in a similar antisymmetric state but with a delocalized position \cite{orthey2019nonlocality}. 

At the exceptional point \( e^{\gamma_\text{ep}} = 1.347... \) in Fig. \ref{fig1} (c) and (g), the joint position probability density \(P(n_1,n_2)\) show non-zero probabilities to measure the two particles at any pair of sites \( (n_1,n_2) \) when the initial two-qubit state is symmetric under exchange. On the other hand, anti-bunching still prevails for the antisymmetric two-qubit state, such that there is only a non-zero probability of measuring the two particles when one of the particle is at \( n_i=2t \) while the other is at \( n_j \neq n_i \). Noisy quantum walks have been argued to be optimal for uniform sampling \cite{maloyer2007decoherence}. With the almost uniform probabilities at the exceptional point, it will be interesting to investigate further how such feature can be utilize for uniform sampling on two random variables or even more with its multiple-particle version. We reserve this for future investigation. In all the gain-loss parameter \(\gamma\) values considered except at the exceptional point, the joint position probability densities are consistent with the expected bosonic and fermionic exchange statistics. 

We see the influence of the symmetry of the states with respect to the particle in the the spatial correlations for different gain-loss parameter values. While quantum-to-classical transition is apparent when gain-loss parameter \(\gamma \) becomes too strong at the exceptional point and beyond, the particle-exchange symmetry of the initial state still manifests in the spatial correlations. 

\begin{figure}
    \centering
    \includegraphics[width=\linewidth]{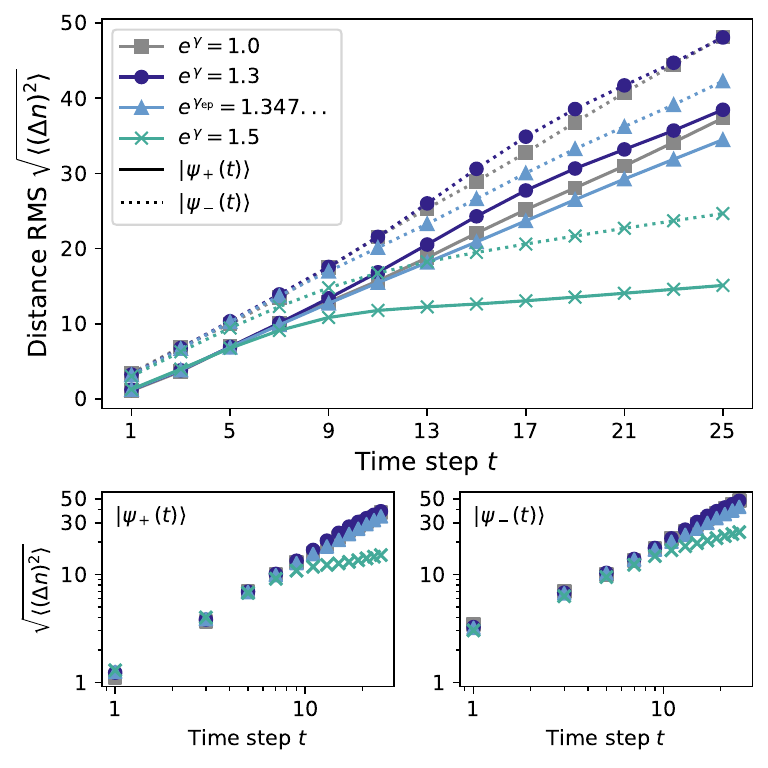}
    \caption{Distance root-mean-squared (RMS) \(\sqrt{\langle ( \Delta n)^2 \rangle }, \Delta  n = n_1-n_2 \) for the non-unitary quantum walks with model parameters \( \theta_1=\pi/4, \theta_2=-\pi/7, \phi=0 \) and varying gain-loss parameter \(\gamma\). The bottom row shows \( \sqrt{\langle ( \Delta n)^2 \rangle } \) in log-log scale. For \(e^\gamma \leq e^{\gamma_\text{ep}}\), \(\sqrt{\langle ( \Delta n)^2 \rangle } \) grows linearly. On the other hand, when gain-loss is sufficiently strong with  \(e^\gamma > e^{\gamma_\text{ep}}\), \(\sqrt{\langle ( \Delta n)^2 \rangle } \) grows sub-linearly after some steps. }
    \label{fig2}
\end{figure}

\subsection{Average distance between the two particles}

The distance between the two particles,  \(\Delta n = n_1-n_2 \), at any instant of time will have a zero average value due to particle exchange symmetry. To further characterize the spatial correlations, we consider the root-mean-squared (RMS) of the distance \(\Delta n \) for the two initial states \(|\psi_\pm \rangle \) with varying values of the gain-loss parameter \(\gamma\). From the joint position probability density \(P(n_1,n_2,t)\), we calculate for the distance RMS by taking the square-root of
\begin{equation}
    \langle \big(\Delta n(t)\big)^2 \rangle = \sum_{n_1,n_2} P(n_1,n_2,t) \; (n_1-n_2)^2.
\end{equation}
Fig. \ref{fig2} shows that the distance RMS \( \sqrt{\langle \big(\Delta n(t)\big)^2 \rangle} \) is consistently larger in the antisymmetric state \( |\psi_-\rangle \) than that of the symmetric state \( |\psi_+\rangle \) at any gain-loss parameter \( \gamma \). This reflects the effect of the exchange symmetry in the dynamics of the two particles -- that antisymmetric state \( |\psi_-\rangle \) tend to anti-bunch leading to a larger distance RMS \( \sqrt{\langle (\Delta n )^2 \rangle} \). 

In the unbroken PT-symmetry phase (\(e^\gamma  < e^{\gamma_\text{ep}} \)), we see that the average distance between the two particles scales linearly in time, i.e. \( \langle (\Delta n)^2 \rangle \sim t \). This is also observed in unitary quantum walks \cite{omar2006quantum}. Since \(P(n_1,n_2,t)\) is maximal around \( (n_i=\pm 2t, n_j =-n_i) \), then the distance when the two particles are at \( (n_i=\pm 2t, n_j =-n_i) \) leads the sum \( \sum (\Delta n)^2 = (4t)^2 + \cdots\) . Thus, \( \sqrt{\langle (\Delta n)^2 \rangle} \sim \sqrt{(4t)^2} \sim t\). This can be simply attributed to the non-interaction of the particles, and hence, the one-particle wavefunction still spreads linearly in time step with maximum amplitudes at either or both ends of the wavefunction.

At the exceptional point \( e^{\gamma_\text{ep}} \), the distance RMS  \(\sqrt{\langle (\Delta n )^2 \rangle} \) also scales linearly in time. Consider the symmetric state \(|\psi_+\rangle \) in Fig. \ref{fig1}-(c) in which its symmetry allows non-zero probabilities  \(P(n_1,n_2,t)\) at all possible pair of sites. Equivalently, we can write
\begin{equation}
   \langle (\Delta n)^2 \rangle = \sum_{x=0,2,...}^{4t} P(x) \; x^2
\end{equation}
where \(P(x)\) is the probability that the two particles is with distance \(x\). We can approximate \( P(x) \) as uniform at all \(x\) which is reasonable especially when \(t\) is large. This gives \(P(x) = 1/(2t+1) \sim 1/t\). Now, the partial sum of squares gives \(\sum_{0,1,...}^{2t} (2x)^2 = [4t(4t+1)(4t+1)]/3 \sim t^3 \). Hence, \( \langle (\Delta n)^2 \rangle \sim t^2 \) and \( \sqrt{\langle (\Delta n)^2 \rangle} \sim t \). Similar argument can be made for the antisymmetric state \(|\psi_-\rangle \) but note that its antisymmetry with respect to particle exchange leads to nonzero probabilities only at distinct sites \(n_i \neq n_j\). Fig. \ref{fig2} also shows the distance RMS \( \sqrt{\langle (\Delta n)^2 \rangle}\) in the log-log scale which supports the linear scaling of the distance RMS in the unbroken PT-symmetry phase and at the exceptional point.

In the broken PT-symmetry phase \(e^\gamma  > e^{\gamma_\text{ep}} \), the distance RMS \( \sqrt{\langle (\Delta n)^2 \rangle} \) deviates from the linear scaling after a few steps and becomes sublinear for both the symmetric and antisymmetric states \( |\psi_\pm\rangle \). This signify the growing effects of decoherence after few steps. The evaluation of \( \langle n_1,n_2|\varrho_\text{P}(t) |n_1,n_2 \rangle \) in the broken PT-symmetry regions reveals that a Gaussian function with variance \(\sigma^2 \sim t \) dominates \( \langle n_1,n_2| \varrho_\text{P}(t) |n_1,n_2 \rangle \) and that the other terms significantly vanish outside \( n_i = \pm \sqrt{t} \). Since the particles are indistinguishable, and clearly from Eq. (\ref{eq:reduced_density_operator_2position}), the one-particle position probability densities are equal, \(P(n_1,t) = P(n_2,t)\). 
For the symmetric state \(|\psi_+ \rangle \) in Fig. \ref{fig1}-(d), we see that the one-particle position probability density \(P(n_i)\) can be well approximated by a Gaussian function, indicating a measurement results that can be simply explained by a classical random walk. Similarly, the joint position probability density \(P(n_1,n_2,t)\) can be approximated by a product of two Gaussian distributions \( \mathcal{N}(n_i;  \sigma,\bar{n})\) with variance \(\sigma^2 \sim t \) and mean \(\bar{n}\). Now,  \(P(n_1,n_2,t) =  \mathcal{N}(n_1; \sigma,\bar{n})  \mathcal{N}(n_2;\sigma,\bar{n})\), which would produce \(P(n_1,t)=P(n_2,t)\). Using the properties of Gaussian distribution we obtain \( \langle (n_1-n_2)^2 \rangle = 2\sigma^2 \sim t \) and \( \sqrt{ \langle (\Delta n)^2 \rangle } \sim \sqrt{t}\).

For the antisymmetric state, Fig. \ref{fig1}-(h) shows two peaks in the joint position probability density \(P(n_1,n_2,t)\) and that \(P(n,n,t) = 0\) as a consequence of the antisymmetry with respect to the particle exchange. We find that the joint position probability density \(P(n_1,n_2,t)\) can be well approximated by
\begin{eqnarray}
    P(n_1,n_2,t) &\propto& (\mathcal{N}(n_1;\sigma,\bar{n})\mathcal{N}(n_2;\sigma,-\bar{n}) \nonumber \\ && - \mathcal{N}(n_1;\sigma,-\bar{n})\mathcal{N}(n_2;\sigma,\bar{n}) )^2,
\end{eqnarray}
the square ensures positivity while satisfying \(P(n,n,t) = 0\). This would also produce \(P(n_1)=P(n_2)\) such that there are two peaks at \(n_i = \bar{n}\) and \(n_i = -\bar{n}\). Again, using the properties of Gaussian distribution, the distance RMS \( \sqrt{ \langle (\Delta n)^2 \rangle } \sim \sqrt{t}\). Hence, in these non-unitary quantum walks, the quantum-to-classical transition can be observed in the linear to sublinear growth of the average distance between the two particles.  

\section{Entanglement dynamics} \label{sec:sec4}

Entanglement is an important resource for quantum information processing. For instance, as a resource for some quantum algorithms, the qubits will be prepared in a highly entangled state to encode the quantum information \cite{chitambar2019quantum}. In our case, we are interested on how the entanglement between the two particles is affected by dissipation in this model of non-unitary quantum walks, as well as how the particle exchange symmetry will play a role in the entanglement dynamics. From Eq. (\ref{eq:explicit_density_operator}), the two-qubit reduced density operator is 
\begin{eqnarray}
    \rho_\text{Q}^\pm(t) &=& \text{Tr}_\text{P} \big( \rho^\pm(t) \big) \nonumber \\
    &=& \sum_{k_1,k_2}\sum_{q,q'} (\pm 1)^{q\oplus q'}\rho^{q,q'}_{k_1}(t) \otimes \rho^{\bar{q},\bar{q}'}_{k_2}(t) 
    \label{eq:two-qubit-density-operator}
\end{eqnarray}
where \( \rho^{q,q'}_{k_1}(t) =  \rho^{q,q'}_{k_1,k_1}(t) \) after tracing out the position degrees-of-freedom of the two particles. Then, we use the normalized reduced density operator \( \varrho_\text{Q}^\pm(t) \) to obtain a normalized probability distribution. For a qubit, we have \(\varrho_\text{q}^\pm(t) = \text{Tr}_{\text{q}'}\big( \varrho_{\text{Q}}^\pm(t) \big) \) such that \(\text{q} \neq \text{q}'\). The low-dimensionality of the qubit and two-qubit reduced density operators allows us to calculate properties and analyze dissipation-induced decoherence for longer time steps.

\subsection{Entanglement entropy}

The entanglement entropy measures the entanglement between a part of a system with the rest of the system. A zero value implies a separable state and a maximal value implies a maximally entangled state. The entanglement between the two-qubit subspace \( \text{Q} \) and the two-particle position subspace \( \text{P} \) can be characterized by the von Neumann entanglement entropy
\begin{equation}
    E_\text{Q}(t) = -\text{Tr} \big( \varrho_\text{Q}(t) \ln \varrho_\text{Q}(t) \big),
\end{equation}
Similarly, the entanglement entropy between a qubit and the rest of the system \( E_\text{q}(t) = -\text{Tr} \big( \varrho_\text{q}(t) \ln \varrho_\text{q}(t) \big) \). Initially, we have \( E_\text{Q}(0) = 0\) since the two-qubit state is separable from the position state, and \( E_\text{q}(0) = \ln 2 \) since the two qubits are maximally entangled in the symmetric and antisymmetric states.

\begin{figure}
    \centering
    \includegraphics[width=\linewidth]{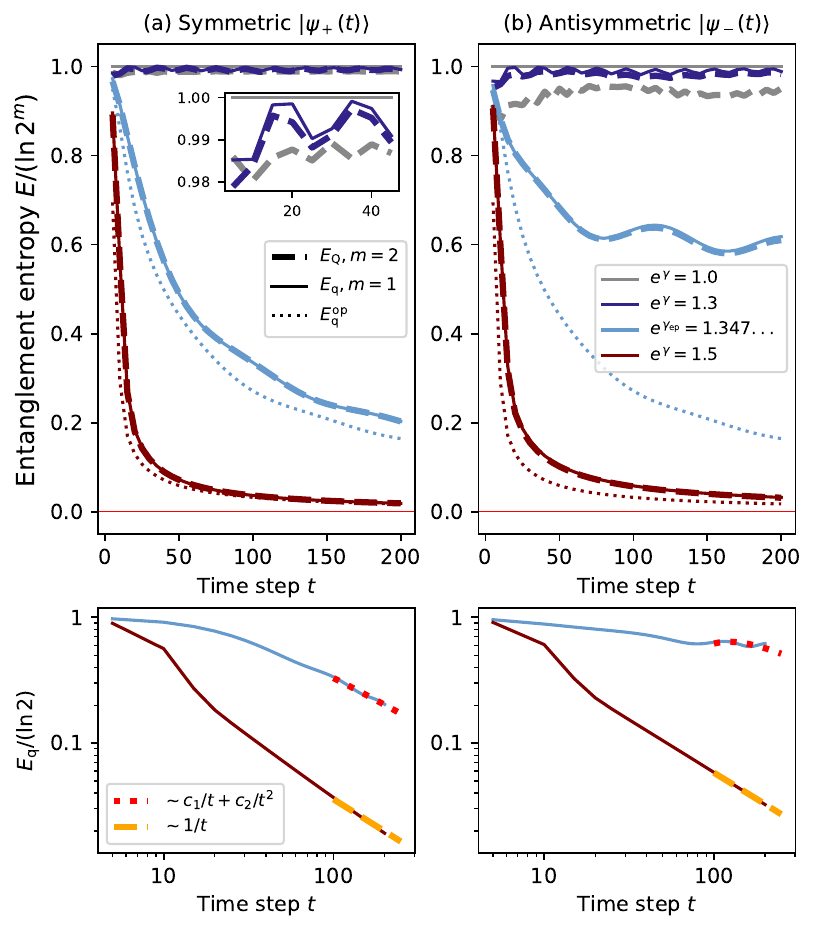}
    \caption{The top row plots the two-qubit and qubit entanglement entropies \( E_{\text{Q}} \) and \( E_\text{q} \) in dashed and solid lines, respectively, for different gain-loss parameter \( \gamma \) values with the other model parameters set at \(\theta_1=-\pi/4, \theta_2=-\pi/7\) and \(\phi=0\). Both  \( E_{\text{Q}} \) and \( E_\text{q} \) remain near maximal in the unbroken PT-symmetry phase  \(e^\gamma < e^{\gamma_\text{ep}}\), while they are decreasing for the quantum walks at the exceptional point and in the broken PT-symmetry phase \(e^\gamma \geq e^{\gamma_\text{ep}}\). At \(e^\gamma_ep\), \( E_\text{q}\) and \(E_\text{Q}\) decay slower in the antisymmetric state \( | \psi_- \rangle \) than in symmetric state \( |\psi_+ \rangle \). The qubit entanglement entropy \(E^\text{op}_\text{q}\) of one-particle quantum walk is included as dotted line which shows \(E^\text{op}_\text{q} < E_\text{q} \) for states \( |\psi_\pm \rangle \). The difference \(E^\text{op}_\text{q} - E_\text{q} \) is larger in \( |\psi_-\rangle \). The bottom row shows the decay of \( E_{\text{q}} \) at the exceptional point and in the broken PT-symmetry as compared with the results of asymptotic analysis on \(\varrho_\text{q}(t)\).}
    \label{fig3}
\end{figure}

In Fig. \ref{fig3}, the qubit entanglement entropy \( E_\text{q}\) remains close to the maximal value through the non-unitary quantum walk in the unbroken PT-symmetry phase. Moreover, the two-qubit subspace Q that is initially uncorrelated with the rest of the system immediately becomes highly entangled with the two-position subspace P as is evident from the growth of two-qubit entanglement entropy \(E_\text{Q}\). These observations hold for both the symmetric and antisymmetric states \( |\psi_\pm \rangle \). At the exceptional point and in the broken PT-symmetry state, we see that the entanglement entropy decays. Interestingly, the qubit and two-qubit entanglement entropies of the antisymmetric state \(|\psi_-\rangle\) decays slower than that of the symmetric state \(|\psi_+ \rangle \) at the exceptional point.

Since the particles are non-interacting, analysis done in Ref. \cite{itable2023entanglement} to study the entanglement dynamics of the one-particle version of these non-unitary quantum walks can be adapted here. The eigenvalues of the modal evolution operator \(U_{k_i}\) are \(e^{\mp \mathrm{i} \epsilon_k}\). In the qubit and two-qubit reduced density operators \(\varrho_\text{q}(t)\) and \(\varrho_\text{Q}(t)\), the time-dependence would originate from the factors \(e^{\mp 2\mathrm{i}\epsilon_k t}\). In the unbroken PT-symmetry phase the quasi-energies \(\pm \epsilon_k \) phase are entirely real, a stationary phase method analysis predicts that the oscillations in \(E_\text{Q}(t)\) and \(E_\text{q}(t)\) eventually cancel out in the infinite-time limit and these quantities converge to constant values.

At the exceptional points, the quasi-energies have gap closing at certain coalescing modes. A coalescing mode would have a singular contribution to the density operator that eventually leads to the vanishing of the entanglement. A Jordan decomposition on the coalescing mode predicts that the qubit entanglement entropy \(E_\text{q}(t)\) vanishes with
\begin{eqnarray}
    E_\text{q}(t \gg 1) &\sim& -\big(1-p(t)\big)\ln  \big(1-p(t)\big) -  p(t) \ln  p(t)  , \qquad \nonumber \\
    && \qquad \; p(t) = c_1/t + c_2/t^2
\end{eqnarray}
 where \(c_j\)'s are parameter-dependent constants. On the other hand, in the broken PT-symmetry phase the imaginary quasi-energies make the dynamical factors \(e^{\mp 2\mathrm{i}\epsilon_k t}\) exponentially increase in time. In the infinite-time limit, a product state is reached between a dominant mode with largest imaginary energy and a parameter-dependent qubit state. A steepest descent method predicts that the entanglement entropy \(E_\text{q}( t \gg 1)\) vanishes such that \(p(t) \sim 1/t\). Similar trends can be argued for the two-qubit entanglement entropy \(E_\text{Q}(t)\).  Fig. \ref{fig3} shows how these predictions agree with how the entanglement entropies \(E_\text{q}\) and \(E_\text{Q}\) decay at large times.

 We also consider the entanglement entropy \(E^\text{op}_\text{q} \) of a one-particle quantum walk with initial state \(|\psi_\text{op}(0)\rangle = |0,0\rangle \) at the exceptional point and in the broken PT-symmetry phase. We compare  \(E^\text{op}_\text{q} \) with the entanglement entropy \(E_\text{q}\) of a qubit in the two-particle quantum walk. Fig. \ref{fig3} shows that consistently \(E^\text{op}_\text{q} < E_\text{q}\), the difference is even more pronounced when in the two-particle quantum walk the initial state is in the antisymmetric state \(|\psi_-\rangle\). Thus, the addition of another qubit for the two-particle quantum walk and initializing the two qubits in the antisymmetric state enable a qubit to maintain some entanglement with the rest of the system for longer times. This feature can be explored further as a potential model for physical qubits that are more tolerant against dissipation.

 \begin{figure}
    \centering
    \includegraphics[width=0.8\linewidth]{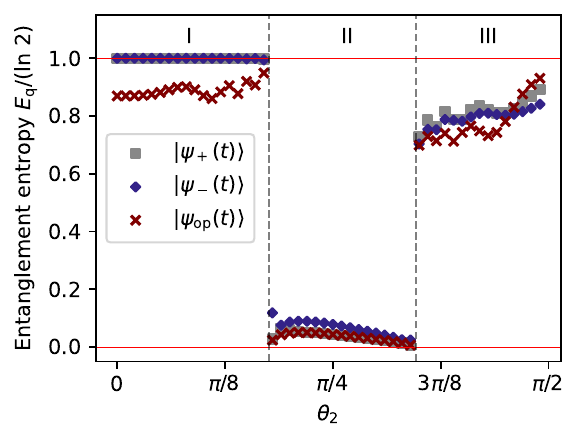}
    \caption{Qubit entanglement entropy \(E_\text{q}\) at time step \(t=100\) across \(\theta_2\) values representing non-trivial (I) and trivial (III) topological phases in the unbroken PT-symmetry phase, and the broken PT-symmetry phase (II); with the other model parameters set at \(\theta_1=-\pi/4\) and \(\phi=0\). \(E_\text{q}\) of symmetric/antisymmetric state \(|\psi_\pm(t) \rangle \) is more robust than \(E_\text{q}\) of the one-particle version  \(|\psi_\text{op}(t)\rangle \) in the non-trivial topological phase.}
    \label{fig4}
\end{figure}

The behavior of the entanglement entropy \(E^\text{op}_\text{q}\) can distinguish the topological phases in one-particle quantum walks \cite{itable2023entanglement,wang2020robustness}. In the unbroken PT-symmetry phase, there exists a non-trivial topological phase characterized by an entanglement entropy \(E^\text{op}_\text{q}\) that is more robust against variations in the model parameters than in the trivial topological phase. More so, the entanglement entropy \(E^\text{op}_\text{q}\) is higher in the non-trivial topological phase than in the trivial topological phase. As shown in Fig. \ref{fig4}, adding another particle and maximally entangling the two particles in their internal qubit degree-of-freedom initially, the qubit entanglement entropy \(E_\text{q}\) is even more robust against variations in the model parameters in the non-trivial topological phase. Hence, distinguishing between the two topological phases of this quantum walk can be done more accurately by considering two particle but it requires more resources as the two particles need to be prepared in a maximally entangled state. 

\subsection{Genuine multipartite entanglement from concurrence}

We may envision the quantum walk of the two particles to occur on different lines. This begs whether the entanglement initially on the two qubits can lead to entanglement between their positions on their respective lines. Hence, we consider if there exists a genuine multipartite entanglement (GME) in this four-partition system \( (\mathcal{H}_{\text{p}_1} \otimes \mathcal{H}_{\text{q}_1} \otimes \mathcal{H}_{\text{p}_2} \otimes \mathcal{H}_{\text{q}_2} ) \). We consider GME concurrence \cite{orthey2019nonlocality,ma2011measure},
\begin{equation}
    \mathscr{C}_\text{GME}\big( \varrho(t) \big) = \min_{\text{A}_i \in \text{A}} \mathscr{C}(\varrho_{\text{A}_i}(t)) =  \min_{\text{A}_i \in \text{A}} \sqrt{2 \mathcal{L}\big(\varrho_{\text{A}_i}(t) \big)},
\end{equation}
where \( \mathcal{L}(\varrho) = 1-\text{Tr} (\varrho^2) \) is the linear entropy and A is the set of all possible bipartitions. This GME concurrence has been used in a unitary two-particle quantum walk with an initially maximally entangled two-qubit state but with a delocalized position  \cite{orthey2019nonlocality}. There is a genuine multipartite entanglement if the state is nonseparable in all possible bipartitions. There are only four unique bipartitions with the following reduced density operators:  the two-qubit reduced density operator \( \varrho_\text{Q}(t)\), the qubit reduced density operator \(\varrho_\text{q}(t) \), the one-particle reduced density operator \(\varrho_{\text{S}}(t) = \text{Tr}_{\text{S}'}\big( \varrho(t) \big) \), and the position reduced density operator \( \varrho_\text{p}(t) =\text{Tr}_{\text{q}}\big(\varrho_{\text{S}'}(t) \big) \).

\begin{figure}
    \centering
    \includegraphics[width=\linewidth]{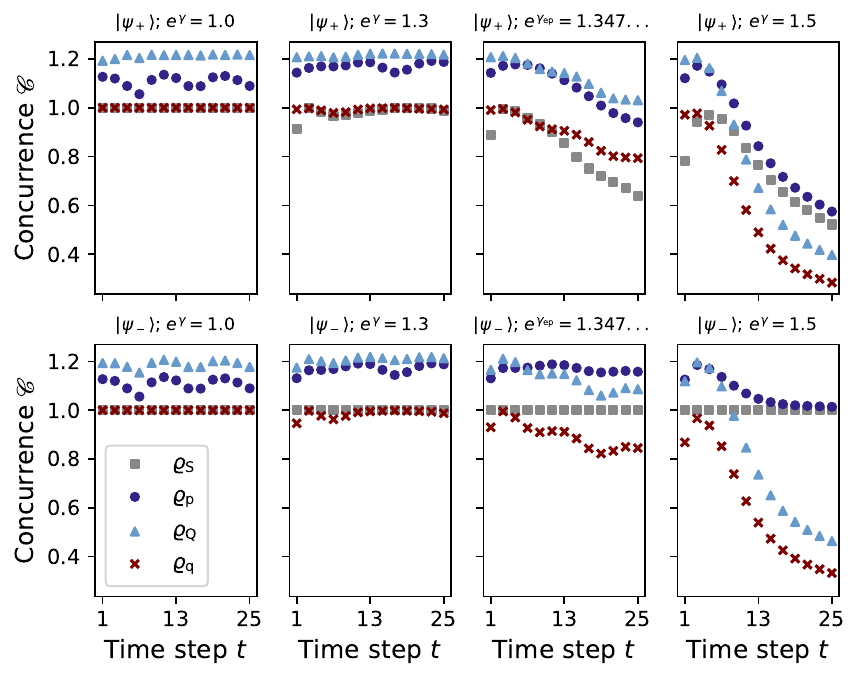}
    \caption{Concurrences \(\mathscr{C} \big(\varrho_\text{A}(t)\big)\) for all the possible unique bipartitions with the reduced density operators obtained for  \(\text{A}_i =\) Q (two-qubit), q (qubit), S (particle  or position+qubit), and p (position), and for different gain-loss parameter value \(\gamma\), while the other model parameters are set at \(\theta_1=-\pi/4, \theta_2=-\pi/7\) and \(\phi=0\). The top and bottom row shows \(\mathscr{C}\) for the symmetric state \(|\psi_+\rangle \) and antisymmetric state  \(|\psi_-\rangle \), respectively. Genuine multipartite entanglement vanishes as the quantum walk evolves at the exceptional point \( e^\gamma = e^{\gamma_\text{ep}} \) and in the broken PT-symmetry phase \(e^\gamma > e^{\gamma_\text{ep}} \). However, the one-particle concurrence \(\mathscr{C}(\varrho_\text{S}) \) in the antisymmetric state is constant and unaffected by dissipation.}
    \label{fig5}
\end{figure}

Fig. \ref{fig5} shows the concurrence \(\mathscr{C}\) in all the unique bipartitions. A \( \mathscr{C} =0 \) indicates the absence of entanglement between the two parts considered in the bipartition. Hence, a decreasing \( \mathscr{C} =0 \) at any of the bipartitions implies a decrease of entanglement in that bipartition and a decrease of genuine multipartite entanglement in the system. Fig. \ref{fig5} shows the presence of genuine multipartite entanglement at the unitary limit and in the unbroken PT-symmetry phase. This could persist in time since we know that the entanglement entropy \( E_\text{Q}\) and \(E_\text{q} \) have nonzero values at the infinite-time limit.  

On the other hand, the GME concurrence \( \mathscr{C}_\text{GME}\) decreases at the exceptional point and in the broken PT-symmetry phase even after a few steps indicating the loss of GME. However, in the antisymmetric state \(|\psi_- \rangle \), it is interesting that the concurrence \(\mathscr{C}\) for the S partition which characterizes the entanglement between the two particles remains constant even in the exceptional point and in the broken PT-symmetry phase. This constant concurrence can be derived explicitly from the reduced density operator \(\rho^\pm_{\text{S}}\),
\begin{equation}
    \rho^\pm_{\text{S}} = \frac{1}{2} \sum_{k_1,k_1',k_2} \sum_{q,q'} (\pm 1)^{q\oplus q'} |k_1\rangle \langle k_1'| \otimes \rho_{k_1,k_1'}^{q,q'} \text{Tr} \big( \rho_{k_2}^{\bar{q},\bar{q}'} \big).
\end{equation}
With the trace of the full density operator \( \rho(t) \) being 
\begin{equation}
    \text{Tr}(\rho^\pm)  = \frac{1}{2} \sum_{k_1,k_2} \sum_{q,q'} (\pm 1)^{q\oplus q'}  \text{Tr}\big(\rho_{k_1}^{q,q'}\big)  \text{Tr} \big( \rho_{k_2}^{\bar{q},\bar{q}'} \big),
\end{equation}
and that
 \begin{eqnarray}
        \text{Tr}\big( (\rho^\pm_{\text{S}})^2 \big) &=& \frac{1}{4} \sum_{q,q'} \sum_{s,s'} (\pm 1)^{q\oplus q' \oplus s \oplus s'}  \nonumber \\
        && \times \bigg( \sum_{\substack{k_1,k_1',\\k_2,k_2'}}  \text{Tr}(\rho_{k_1}^{s',q}) \text{Tr}(\rho^{q',s}_{k_2})  \text{Tr} ( \rho_{k_1'}^{\bar{q},\bar{q}'})  \text{Tr}( \rho_{k_2'}^{\bar{s},\bar{s}'}) \bigg), \nonumber \\
\end{eqnarray}
we find \( \text{Tr} \big( (\rho^-_{\text{S}})^2\big)= 2\big( \text{Tr}(\rho^-)\big)^2 \) for the antisymmetric state at any time \(t\). Hence, after normalization, the linear entropy is \(\mathcal{L}\big(\varrho^-_{\text{S}}(t)\big)=1/2\), and the concurrence is \( \mathscr{C} \big(\varrho^-_{\text{S}}(t)\big) = 1 \). In the broken PT-symmetry phase, we may attribute the constant concurrence to the non-vanishing spatial anti-correlation between the two particles, as shown in Fig. \ref{fig1} (g) and (h). A similar result in regards to the resilience of concurrence against dissipation was also shown in the study of entanglement generation between two spins induced by a noisy magnetic medium \cite{zou2022bell}. The work showed persistent and maximal concurrence can be achieved with proper post-selection at the exceptional point and broken PT-symmetry region of the model. We reserve for future investigations the potential tasks this constant concurrence in the antisymmetric state \(|\psi_-\rangle \) can be further exploited, especially in multi-particle quantum walks. 

\section{Conclusions} \label{sec:sec5}

In conclusion, we investigated the generation of quantum correlations in a two-particle discrete-time non-unitary quantum walk with gain and loss. The model has PT-symmetry due to the alternating and balanced gain and loss. In the unbroken PT-symmetry phase, the dissipation is low enough that the dynamics, while non-unitary, still have quantum coherence. Consequently, we observed here similar features already established in unitary two-particle quantum walks \cite{omar2006quantum}, such as significant probabilities of (anti-)bunching in the (anti-)symmetric state, linear growth of the average interparticle distance, and persistent bipartite entanglement. Furthermore, the two-particle quantum walk showed an enhanced ability of the entanglement entropy to distinguish the non-trivial topological phase from the trivial topological phase compared to the one-particle version.

In the PT-symmetric phase, in which the dissipation becomes strong enough and decoherence dominates the dynamics, the sublinear growth of the average distance of the particles and eventual loss of bipartite entanglement (as measured either through entanglement entropy or concurrence) suggest a transition toward a classical random walk description. However, we found that the one-particle concurrence of the antisymmetric state is unaffected by dissipation. Moreover, at the exceptional point, the entanglement entropy in the antisymmetric state decays significantly slower than that of the symmetric state. These observations motivate further study into why the antisymmetric state appears more resilient to dissipation.

These findings suggest that quantum correlations in such non-unitary dynamics can be strongly influenced by the particle exchange symmetry and may serve as indicators of both quantum-to-classical transitions and the underlying topological phases of the system. Future work may explore how these features can be exploited in quantum information protocols with engineered dissipation \cite{harrington2022engineered,zou2022bell} or extended to multi-particle quantum walks.

\begin{acknowledgments}
GMMI acknowledges the DOST Science and Technology Fellows Program for the general support and the ASTI Quantum Circuit Simulator Project (DOST-GIA Funded Project No. 10873) for the computational resources.
\end{acknowledgments}

\bibliography{bibfile}


\end{document}